\documentclass{PoS}

\title{Finite volume corrections to LECs in Wilson and staggered ChPT }

\ShortTitle{Finite volume corrections to LECs in Wilson and staggered ChPT }

\author{\speaker{Fabrizio Pucci}\\
        Fakult\"{a}t f\"{u}r Physik, Universit\"{a}t Bielefeld, D-33615 Bielefeld, Germany\\
        E-mail: \email{pucci@physik.uni-bielefeld.de}}

\author{Gernot Akemann\\
        Fakult\"{a}t f\"{u}r Physik, Universit\"{a}t Bielefeld, D-33615 Bielefeld, Germany\\
        E-mail: \email{akemann@physik.uni-bielefeld.de }}

\abstract{We study the simultaneous effect of finite volume and finite lattice spacing corrections in the framework of chiral perturbation theory (ChPT) in the epsilon regime, for both the Wilson and staggered formulations. In particular the finite volume corrections to the low energy constants (LECs) in Wilson and staggered ChPT are computed to next-to-leading order (NLO) in the $\epsilon-$expansion. For Wilson with $N_f$=2 flavours and staggered with generic $N_f$ the partition function at NLO can be rewritten as the LO partition function with renormalized effective LECs.}

\FullConference{The 30th International Symposium on Lattice Field Theory\\
                 June 24 - 29,  2012\\
                 Cairns, Australia}

\begin{document}

\section{Introduction}
\noindent
In the last years the lattice QCD simulations near the chiral limit drive renewed interest to understand this limit with analytical approaches, and indeed a lot of efforts have been done in the study of the low energy effective theories. It is important in the numerical simulation to have the lattice spacing effects under control that also break explicitly chiral symmetry. Wilson and staggered chiral perturbation theory (WChPT and SChPT) provide the framework in which one can study these UV cut-off effects. \newline
In the Wilson and staggered ChPT lagrangians in addition to the continuum Gasser-Leutwyler terms \cite{Gasser:1983yg,Weinberg:1978kz} there are additionals $O(a^2)$ contributions, and hence new low energy effective constants (LECs) that need to be introduced. More in detail for leading order (LO) WChPT with two flavors only one new LEC enters, usually denoted with $c_2$ \cite{Sharpe:1998xm,Aoki:2003yv}, while for SChPT six new LECs $C_i$ need to be introduced \cite{LeeSharpe,AubinBernard}.\newline
Here we study these theories in a finite volume in the so-called $\epsilon$-regime \cite{Gasser:1986vb,Gasser:1987ah}, namely when the pion Compton wave length is bigger then the lattice size $L$
\begin{equation}m_{\pi} L \ll 1 . \end{equation}
\noindent
This regime is extremely intriguing since systematic analytical calculations are possible. In particular it has been shown that for both formulations \cite{Damgaard:2010cz,Akemann:2010em,Osborn:2010eq} at LO in the $\epsilon$-expansion these theories are equivalent respectively to Wilson chiral Random Matrix Theory (WChRMT) for the Wilson formulation and to staggered Chiral RMT for the staggered one.\newline
In these chiral theories one has also to understand the relative size between the quark mass $m$ and the lattice spacing $a$. For example in WChPT there are three possible power countings
\cite{Shindler:2009ri,Bar:2008th}
that are usually applied depending on the appearance of the cut-off effect at the LO, the so-called Aoki-regime, at Next-to-Leading order (NLO) called GSM$^*$ regime, or
Next-to-Next-to-Leading order (NNLO) called GSM regime in which NLO corrections to the spectral density of the Wilson Dirac operator have already been computed \cite{Necco:2011vx}.
Through these proceedings we analyze WChPT and SChPT using the first power counting scheme, namely when
 \begin{equation}m \sim a^2 \Lambda^3_{QCD} \sim O( \epsilon^4 )\ .\end{equation}
\noindent
It is usually known also as large cut-off effect regime since at LO both the mass and the cut-off terms contribute with the same strength to the chiral symmetry breaking.\newline
In particular we address the problem of the extension up to $O(\epsilon^2)$ of the partition function for WChPT with $N_f$=2 flavors and for SChPT for generic $N_f$. The main result that we present here and in a forthcoming publication \cite{Akemann:2012bc} is the possibility to write the NLO partition function for both theories as
the LO one with renormalized LECs. This is analogous to what happens in continuum chiral perturbation theory \cite{Damgaard:2001js,Damgaard:2007xg,Akemann:2008vp} with the only difference that here the renormalization factor of the LECs can not be written in terms of the geometric data of the lattice alone. This result opens up the possibility to extend the relations WChPT/WChRMT
and SChPT/SChRMT up to NLO.\newline

\section{Wilson Chiral Perturbation Theory for $N_f$ = 2}
\noindent
The Wilson chiral lagrangian for the two-flavor case with degenerate mass $m$ can be written as
%\small
\begin{equation}
L_{LO} =\frac{F^2}{4} Tr\left[ \partial_{\mu} U \partial_{\mu} U^\dag \right] - \frac{\Sigma}{2}  Tr\left[ M^{\dagger} U + U^{\dagger} M \right]
  + a^2 c_2\,\left(Tr\left[ U + U^{\dagger}\right]\right)^2 \ .\label{actionnf2}
\end{equation}
%\normalsize
\noindent
As usual $\Sigma$ is the chiral condensate, $F$ is the pion decay constant and $c_2$ is the new LEC of WChPT. The main idea underlying the construction of the theory in the $\epsilon$-regime is that, since the zero modes dominate the path integral, one has to
threat them non-perturbatively, in contrast to the propagating quantum fluctuations. Thus the usual parameterization for the matrix $U$ is given by
\begin{equation}
U(x) = U_0\, \, exp\left[i \frac{\sqrt{2}}{F}\, \xi(x) \right]\ ,
\label{Upara}
\end{equation}
\noindent
where $U_0$ is the two by two unitary matrix describing the zero-modes and $\xi$ are the fluctuations. Since also the NLO lagrangian needs to be considered in the present calculation, following \cite{Hansen:2011mc,Bar:2003mh} we write it as
%\small
\begin{eqnarray}
  L_{ NLO} &=&  a\, c_0\,
Tr\left[ \partial_{\mu} U\, \partial_{\mu} U^{\dagger}\right]
Tr\left[ U +  U^{\dagger}\right]
+ a m\, c_3\, \left(Tr\left[ U +  U^{\dagger}\right]\right)^2 \nonumber\\
&&+ a^3 d_1 \, Tr\left[ U +  U^{\dagger}\right] + a^3 d_2 \left(Tr\left[ U +  U^{\dagger}\right]\right)^3\ , \label{actionnf2NLO}
\end{eqnarray}
%\normalsize
\noindent
where 4 new LECs are introduced, namely $c_0, c_3, d_1$ and $d_2$.  The idea is to expand the action
\begin{equation}S = \int d^4x\, \left( L_{ LO} + L_{ NLO} \right)\end{equation}
\noindent
up to $O(\epsilon^2)$ using the Aoki regime power counting
\begin{equation}
V \sim \epsilon^{-4},\, m\sim \epsilon^4,\, \partial \sim \epsilon, \, \xi(x)\sim \epsilon,\, \, a\sim \epsilon^2. \label{powercounting}
\end{equation}
\noindent
At LO the different contributions can be rearranged as
%\small
\begin{eqnarray}
S^{(0)} &=& \frac{1}{2}\, \int d^4x\, Tr\left[\partial_{\mu} \xi(x) \partial_{\mu} \xi(x)\right]
-  \frac{1}{2}\, m\, V \Sigma\, Tr\left[ U_0 + U_0^{\dagger} \right]
+ a^2 V c_2\, \left(Tr\left[ U_0 + U_0^{\dagger}\right]\right)^2 \\
&\equiv& S^{(0)}_{\partial^2}+S^{(0)}_{U_{0}}.
\end{eqnarray}
%\normalsize
Now the trick is to rewrite the partition function by separating the integration over the zero-modes from the integration over the Gaussian fluctuations as
\begin{equation}Z = \int_{SU(2)} [d_H U(x)]\, e^{-S} = \int_{SU(2)} d_H U_0\, \, e^{-S_{U_0}^{(0)}}\, \, Z_{\xi}(U_0)\end{equation}
\noindent
with
\begin{equation}Z_{\xi}(U_0) = \int [d\xi(x)]\, J(\xi(x))\,  e^{S_{U_0}^{(0)} - S}.\end{equation}
\noindent
The factor $J(\xi(x))$ is the Jacobian arising from the change of integration variables. At this point we can expand
the function $Z_{\xi}(U_0)$ up to $O(\epsilon^2)$
and then perform all the Gaussian integrals using the expression
%\small
\begin{equation}\int [d\xi(x)]
\exp\left[-S_{\partial^2}^{(0)}\right]\
%e^{- \frac{1}{2}\, \int d^4x\, Tr\left(\partial_{\mu} \xi(x) %\partial_{\mu} \xi(x)\right) }
\xi(x)_{ij}\xi(y)_{kl} = \left( \delta_{il}\delta_{jk} - \frac{1}{N_f}\, \delta_{ij}\delta_{kl}\right)\Delta(x-y)\label{propagator}\end{equation}
%\normalsize
\noindent
in terms of the propagator $\Delta(x-y)$. We easily find that
%\small
\begin{eqnarray}
\nonumber
Z_{\xi}(U_0) &=& N \left\{ 1+\left(
-\, \frac{3\, m V \Sigma}{4 F^2} \Delta(0)
%+\frac{6ac_0+3ac_1}{F^2}
- a^3 d_1 V  \right)
 Tr\left[ U_0 + U_0^{\dagger} \right]
 \right. \\
&& \left.
+ \left(\frac{4 a^2 c_2 V}{ F^2}  \Delta(0)- a m c_3V \right)\left(Tr\left[ U_0 +  U_0^{\dagger}\right]\right)^2  -a^3 d_2 V\left(Tr\left[ U_0 +  U_0^{\dagger}\right]\right)^3\right\}\
%\nn\\&&
\label{32}\end{eqnarray}
\noindent
%\normalsize
where $N$ is an overall normalization. In dimensional regularization the propagator $\Delta(0)$ is finite and can be written as
\begin{equation}\Delta(0) = - \frac{\beta_1}{\sqrt{V}}\end{equation}
\noindent
with $\beta_1$ a numerical coefficient that encodes the geometrical data of the lattice.\newline
Now with some algebraic manipulations, using the properties of the $SU(2)$ group and the relation
%\small
\begin{equation} \left(\frac{3}{2} + 16 \hat{a}^2 c_2\right) \langle  Tr\left[ U_0 + U_0^{\dagger}\right]\rangle + \frac{\hat{m}}{4} \langle \left(Tr\left[ U_0 + U_0^{\dagger }\right]\right)^2\rangle  - \hat{a}^2 c_2 \langle  \left(Tr\left[ U_0 + U_0^{\dagger}\right]\right)^3\rangle - 4 \hat{m} = 0\label{SecondaSU2}\end{equation}
\noindent
%\normalsize
it is immediate to see that re-exponentiating all the terms of the previous expansion, the partition function is equal to the LO one if we use instead of $\Sigma$ and $c_{2}$, the renormalized low energy constants $\Sigma^{ef\!f}$ and $c_{2}^{ef\!f}$ defined as
%\small
\begin{equation}\Sigma^{ef\!f}
%= \Sigma \left( 1- \frac{3}{2 F^2}\Delta(0)\right)
%%+2a\frac{6c_0+3c_1}{mVF^2}
%- \frac{ 2 a^3 d_1 }{m} - \frac{32 a^3 d_2 }{m } + \frac{3 a d_2 }{ m c_2 V}
=\Sigma \left( 1
- \frac{3}{2 F^2}\Delta(0)-\frac{\hat{a}}{\hat{m}\sqrt{V}}
\left(2\hat{a}^2d_1+32\hat{a}^2d_2-3\frac{d_2}{c_2}\right)\right)
\end{equation}
\noindent
and
\begin{equation}c_{2}^{ef\!f}
%= c_2 \left( 1 -   \frac{4}{ F^2}\Delta(0)\right) +  \frac{m\,  c_3}{a}
%+  \frac{m\, \Sigma\, d_2}{4\, c_2\, a}
= c_2 \left( 1 -   \frac{4}{ F^2}\Delta(0)\right)+\frac{\hat{m}}{\hat{a}}
\left(\frac{c_3}{\Sigma}+\frac{d_2}{4c_2}\right)\frac{1}{\sqrt{V}}
. \end{equation}
\noindent
%\normalsize
Here we have defined
\begin{equation}
\hat{m}\equiv m\Sigma V\ \ \mbox{and} \ \ \hat{a}^2\equiv a^2V
\end{equation}
which are of order 1. Thus the NLO partition function reads as
%\small
\begin{eqnarray}
  Z_{NLO} &=& N' \int_{SU(2)} d_H U_0\, exp\left[  \frac{m \Sigma^{ef\!f} V}{2} Tr\left[ U_0 + U_0^{\dagger} \right]
  - a^2\, c_{2}^{ef\!f}\, V\left(Tr\left[ U_0 + U_0^{\dagger}\right]\right)^2\right] \nonumber \\
&=&\frac{N'  }{N}
 Z_{LO}(\Sigma^{ef\!f},c_{2}^{ef\!f}).
\label{ZNLOnf2}
\end{eqnarray}
\noindent
%\normalsize
Since effective LECs given above at NLO depend in a non trivial way on the additional LECs and not only on the geometrical data of the lattice, in principle it is possible to use a finite volume scaling analysis to extract the numerical value of these undetermined NLO LECs. Performing the simulations at two different lattice volume $V_1$ and $V_2$ with geometries
$\beta_1$ and $\beta_2$, WChPT predicts a scaling of the LECs as
\begin{eqnarray}\frac{\Sigma^{ef\!f}(V_1)}{\Sigma^{ef\!f}(V_2)}  &=& 1
+ \frac{3}{2 F^2} \frac{(\beta_1 \sqrt{V_2}- \beta_2 \sqrt{V_1}) }{\sqrt{V_1 \, V_2}}
+ \left(\frac{3 a d_2 }{m c_2 \Sigma}
%-\frac{12ac_0+6ac_1}{m\Sigma F^2}
\right)
\left( \frac{1}{V_1} - \frac{1}{V_2}\right),\\
%\end{equation}
%\begin{equation}
\frac{c_{2}^{ef\!f}(V_1)}{c_{2}^{ef\!f}(V_2)} &=& 1 + \frac{4}{F^2} \frac{(\beta_1 \sqrt{V_2}- \beta_2 \sqrt{V_1} )}{\sqrt{V_1 \, V_2}}.
\end{eqnarray}

\section{Staggered Chiral Perturbation Theory}
\normalsize
\noindent
The staggered chiral lagrangian has been introduced in \cite{LeeSharpe} and \cite{AubinBernard} respectively for the one-flavor theory and for the general $N_f$ case and reads as
%\footnotesize
%\small
\begin{center}
\begin{eqnarray}
\nonumber  L_{LO} &=& \frac{F^2}{8} Tr\left( \partial_{\mu} U \partial_{\mu} U^{\dagger} \right) - \frac{\Sigma}{4} Tr\left( M^{\dagger} U + U^{\dagger} M \right) - a^2 C_1 Tr\left( U \gamma_5 U^{\dagger\,} \gamma_5\right)- a^2 C_6\, \sum_{\mu < \nu} Tr\left( U \gamma_{\mu \nu } U^{\dagger} \gamma_{\mu \nu }\right)\\
\nonumber  &&- a^2 \frac{C_3}{2} \sum_{\mu}\, \left[ Tr\left( U \gamma_{\mu} U \gamma_{\mu}\right) + h.c. \right] - a^2 \frac{C_4}{2}\, \sum_{\mu}\left[  Tr\left( U \gamma_{\mu 5 } U \gamma_{\mu 5}\right) + h.c.\right]\\
\nonumber &&- a^2 \frac{C_{2V}}{4}\, \sum_{\mu} \left[  Tr\left( U \gamma_{\mu}\right) Tr\left( U \gamma_{\mu }\right) + h.c. \right]
- a^2 \frac{C_{2A}}{4}\, \sum_{\mu} \left[  Tr\left( U \gamma_{\mu 5}\right) Tr\left( U \gamma_{\mu 5 }\right) + h.c. \right] \\
% \nonumber
 &&- a^2 \frac{C_{5V}}{4}\, \sum_{\mu} \left[  Tr\left( U \gamma_{\mu}\right) Tr\left( U^{\dagger} \gamma_{\mu }\right) \right]-  a^2 \frac{C_{5A}}{4}\, \sum_{\mu} \left[  Tr\left( U \gamma_{\mu 5}\right) Tr\left( U^{\dagger} \gamma_{\mu 5 }\right) \right].
\label{no}\end{eqnarray}
\end{center}
%\normalsize
\noindent
The $4 N_f \times 4 N_f$ unitary matrix $U$ is parameterized as
%\small
\begin{center} $U = \left(\begin{array}{cccc}
  u & \pi^{+} & K^{+} & ...\\
  \pi^{-} & d & K^0 & ...\\
  K^{-} & \bar{K}^0 & s & ...\\
  ...   &   ...  & ...  & \ddots
\end{array}\right)$
 \end{center}
%\normalsize
\noindent
with  $u,\, \pi^+,\, K ^+\, ...\, $  being $4 \times 4$ matrices that take into account the four taste degrees of freedom (see \cite{AubinBernard} for details). Indeed in the staggered formulations a single staggered Dirac matrix yields four quark tastes due to the doubling problem. These states are degenerate in the continuum but split at non-zero lattice spacing, and as consequence of this breaking new terms and new LECs usually denoted as $C_1, C_{2\, A}, C_{2\, V}, C_3, C_4, C_{5\, A}, C_{5\, V}, C_6$ are introduced in the chiral lagrangian.\newline
In the Aoki regime the lagrangian (\ref{no}) describes the LO unrooted theory. If we want to go beyond we have to consider also the NLO terms that potentially arise from the discretization. However, in contrast to the Wilson theory here the first correction appear only at NNLO, making the computation easily respect to the Wilson case. Thus if we want to study the finite volume correction to the LECs $C_i$ we will have only to expand the LO lagrangian up to the $O(\epsilon^2)$ order.\newline
In order to begin we rewrite the partition function as
\begin{equation}Z = \int_{SU(4 N_f)} [d_H U(x)]\, e^{-S} = \int_{SU(4 N_f)} D_H U_0\, \, e^{-S_{U_0}^{(0)}}\, \, Z_{\xi}(U_0)\end{equation}
\noindent
where as in the previous section we have separated  the integration over the zero-modes $U_0$ from the integration over the fluctuations $\xi$.
Now we can expand the function $Z_{\xi}(U_0)$ up to order $O(\epsilon^2)$, perform the Gaussian integrals over the fluctuations  and finally after re-exponentiating all the terms we can absorbs the $O(\epsilon^2)$ corrections in the renormalized LECs. At the end we can write
\begin{equation}
\nonumber  Z_{NLO} = \frac{N'  }{N} Z_{LO}\left(\Sigma^{ef\!f}, C_{i}^{ef\!f}\right)
\end{equation}
\noindent
where the value of the $\Sigma^{ef\!f}$ and $C_{i}^{ef\!f}$ are given in the following table.
\vspace{0.2cm}
\scriptsize
\begin{center}
\begin{tabular}{|c|c|}
  \hline
     &         \\
  \begin{math}\Sigma^{ef\!f} =  \Sigma \left( 1 - \frac{16 N_f^2 - 1 }{4 F^2 N_f} \Delta(0)\right)\end{math} & $C_1^{ef\!f}= C_1 \left( 1 - \frac{8 N_f}{F^2} \Delta(0)\right)$ \\
          &       \\
  $C_{2V}^{ef\!f}=  C_{2V} - \frac{ C_{2V}(16 N_f^2 -2)+ 4 C_{3} N_f}{2 N_f\, F^2}\, \Delta(0)$  &
  $C_{2A}^{ef\!f}= C_{2A} - \frac{C_{2A}(16 N_f^2 -2)+ 4 C_{4} N_f}{2 N_f\, F^2}\, \Delta(0)$ \\
                &        \\
  $C_3^{ef\!f}= C_3  -  \frac{C_3(16 N_f^2 -2)+ 2\, C_{2V} N_f}{2 N_f\, F^2}\, \Delta(0) $ &
   $C_4^{ef\!f}= C_4 -  \frac{C_4(16 N_f^2 -2)+ 2\, C_{2A} N_f}{2 N_f\, F^2}\, \Delta(0) $ \\
             &              \\
   $C_{5V}^{ef\!f}= C_{5V} \left( 1 - \frac{8 N_f}{F^2} \Delta(0)\right)$ & $C_{5A}^{ef\!f}= C_{5A} \left( 1 - \frac{8 N_f}{F^2} \Delta(0)\right)$ \\
           &        \\
   $C_6^{ef\!f}= C_6 \left( 1 - \frac{8 N_f }{F^2} \Delta(0)\right)$ &   \\
               &             \\
   \hline
\end{tabular}
\end{center}
\begin{center}
\small
\textbf{Table 1.} The renormalized SChPT LECs.
\end{center}
\vspace{0.2cm}
\normalsize
\noindent
 From a tree level expansion of the NLO chiral lagrangian we can reads the NLO masses of the non-neutral mesons\footnote{ For flavor neutral mesons the situation is more complicated and other terms have to be introduced in the chiral lagrangian.} composed of quark $b$ and $c$
\noindent
\begin{equation}m^2 = \mu ( m_b + m_c ) + a^2 \Delta_{\xi_B}^{NLO}.\end{equation}
\noindent
Here the taste splittings $\Delta_{\xi_B}^{NLO}$ depend obviously by the taste state identified by the taste matrix $\xi_B$. All the states fall into 5 different classes : the Pseudoscalar (PS), Axial-Vector (AV), Tensor (T), Vector (V) and Singlet (S) sector. In such channels the taste splitting \cite{AubinBernard} can be written at NLO when inserting our results from table 1.:

\footnotesize
\begin{equation}\Delta_{PS}^{NLO} = 0\end{equation}

\begin{equation} \Delta_A^{NLO} = \frac{16}{F^2} \left( C_1 + 3 C_3 + C_4 + 3 C_6 \right) -  \frac{16}{F^4} \left( 8 N_f [ C_1 + 3 C_6] + \frac{[C_4 + 3 C_3](16 N_f^2 -2)+ 2\, [ 3 C_{2V} + C_{2A} ] N_f}{2 N_f\,}\right)\Delta(0)\end{equation}

\begin{equation}\Delta_T^{NLO} = \frac{16}{F^2} \left( 2 C_3 + 2 C_4 + 4 C_6 \right) - \frac{16}{F^4} \left( 32 N_f C_6 +  \frac{[C_3 + C_4](16 N_f^2 -2)+ 2\, [ C_{2V} + C_{2A} ] N_f}{N_f\,} \right)\Delta(0)\end{equation}

\begin{equation}\Delta_V^{NLO} = \frac{16}{F^2} \left( C_1 + C_3 + 3 C_4 + 3 C_6 \right) -  \frac{16}{F^4} \left( 8 N_f [ C_1 + 3 C_6] + \frac{[3 C_4 + C_3](16 N_f^2 -2)+ 2\, [ C_{2V} + 3 C_{2A} ] N_f}{2 N_f\,}\right)\Delta(0)\end{equation}

\begin{equation}\Delta_I^{NLO} = \frac{16}{F^2} \left( 4 C_3 + 4 C_4 \right) - \frac{32}{F^4} \left(  \frac{[ C_3 + C_4 ](16 N_f^2 -2)+ 2\, [ C_{2V} + C_{2A} ] N_f}{ N_f\,} \right)\Delta(0)\ .\end{equation}
\normalsize

\section{Summary and Discussion}
\noindent
Throughout this paper we have shown that in the $\epsilon$-regime for two-flavor Wilson chiral perturbation theory and for general $N_f$ staggered ChPT the NLO order partition function can be written as the LO one with renormalized effective LECs. \newline
This result leads to several consequences that we will expand upon in a forthcoming publication \cite{Akemann:2012bc}. The first regards the possibility to extend the relations between LO WChPT and SChPT with Chiral Random Matrix Theory in its Wilson and staggered version respectively.\newline
The second consequence is that in WChPT, due to the fact that the finite volume corrections change the mean field potential, this effect  changes the phase boundaries of the theory.\newline
A further point is the extension of our result to  Wilson ChPT with general $N_f$. In that case the situation is more involved since 3 LO and 9 NLO LECs have to be introduced and naturally the chiral lagrangian becomes more involved, including the question of possible constraints on the signs of individual LECs and combinations of these.

\vspace{0.25cm}

\noindent
\textbf{ Acknowledgments}

\noindent
Partial support by the SFB$|$TR12 ``Symmetries and Universality
in Mesoscopic Systems'' of the German research council DFG is acknowledged (G.A.).
F.P. is supported by the Research Executive Agency (REA)
of the European Union under Grant Agreement PITNGA-
2009-238353 (ITN STRONGnet).


\vspace{0.25cm}

\begin{thebibliography}{99}


\bibitem{Gasser:1983yg}
  J.~Gasser and H.~Leutwyler,
  %``Chiral Perturbation Theory to One Loop,''
  Annals Phys.\  {\bf 158} (1984) 142.

\bibitem{Weinberg:1978kz}
  S.~Weinberg,
  %``Phenomenological Lagrangians,''
  Physica A {\bf 96} (1979) 327.

\bibitem{Sharpe:1998xm}
  S.~R.~Sharpe and R.~L.~Singleton, Jr,
  %``Spontaneous flavor and parity breaking with Wilson fermions,''
  Phys.\ Rev.\ D {\bf 58} (1998) 074501
  [hep-lat/9804028].

\bibitem{Aoki:2003yv}
  S.~Aoki,
  %``Chiral perturbation theory with Wilson-type fermions including a**2
  %effects: N(f) = 2 degenerate case,''
  Phys.\ Rev.\  D {\bf 68} (2003) 054508
  [arXiv:hep-lat/0306027].



\bibitem{LeeSharpe}
  W.~J.~Lee and S.~R.~Sharpe,
%  "Partial Flavor Symmetry Restoration for Chiral Staggered Fermions,"
  Phys.\ Rev.\  D {\bf 60} (1999) 114503
  [arXiv:hep-lat/9905023].

\bibitem{AubinBernard}
  C.~Aubin and C.~Bernard,
  %"Pion and Kaon masses in Staggered Chiral Perturbation Theory,"
  Phys.\ Rev.\  D {\bf 68} (2003) 034014
  [arXiv:hep-lat/0304014].

\bibitem{Gasser:1986vb}
 J.~Gasser and H.~Leutwyler,
 %``Light Quarks at Low Temperatures,''
 Phys.\ Lett.\ B {\bf 184} (1987) 83.

\bibitem{Gasser:1987ah}
 J.~Gasser and H.~Leutwyler,
 %``Thermodynamics of Chiral Symmetry,''
 Phys.\ Lett.\ B {\bf 188} (1987) 477.



\bibitem{Damgaard:2010cz}
  P.~H.~Damgaard, K.~Splittorff and J.~J.~M.~Verbaarschot,
  %``Microscopic Spectrum of the Wilson Dirac Operator,''
  Phys.\ Rev.\ Lett.\  {\bf 105} (2010) 162002
  [arXiv:1001.2937 [hep-th]].


 \bibitem{Akemann:2010em}
  G.~Akemann, P.~H.~Damgaard, K.~Splittorff and J.~J.~M.~Verbaarschot,
  %``Spectrum of the Wilson Dirac Operator at Finite Lattice Spacings,''
  Phys.\ Rev.\  D {\bf 83} (2011) 085014
  [arXiv:1012.0752 [hep-lat]].

\bibitem{Osborn:2010eq}
  J.~C.~Osborn,
  %``Staggered chiral random matrix theory,''
  Phys.\ Rev.\  D {\bf 83} (2011) 034505
  [arXiv:1012.4837 [hep-lat]].

%\cite{Shindler:2009ri}
\bibitem{Shindler:2009ri}
  A.~Shindler,
  %``Observations on the Wilson fermions in the epsilon regime,''
  Phys.\ Lett.\ B {\bf 672} (2009) 82
  [arXiv:0812.2251 [hep-lat]].
  %%CITATION = ARXIV:0812.2251;%%

%\cite{Bar:2008th}
\bibitem{Bar:2008th}
  O.~Bar, S.~Necco and S.~Schaefer,
  %``The Epsilon regime with Wilson fermions,''
  JHEP {\bf 0903} (2009) 006
  [arXiv:0812.2403 [hep-lat]].
  %%CITATION = ARXIV:0812.2403;%%

 %\cite{Necco:2011vx}
\bibitem{Necco:2011vx}
  S.~Necco and A.~Shindler,
  %``Spectral density of the Hermitean Wilson Dirac operator: a NLO computation in chiral perturbation theory,''
  JHEP {\bf 1104} (2011) 031
  [arXiv:1101.1778 [hep-lat]].
  %%CITATION = ARXIV:1101.1778;%%


\bibitem{Akemann:2012bc}
  G.~Akemann and F.~Pucci,
 % "Exploring the Aoki regime,"
 [arXiv:1211.3980 [hep-lat]].


\bibitem{Damgaard:2001js}
  P.~H.~Damgaard, M.~C.~Diamantini, P.~Hernandez and K.~Jansen,
%  "Finite-size scaling of meson propagators,"
  Nucl.\ Phys.\  B {\bf 629} (2002) 445
  [arXiv:hep-lat/0112016].

%\cite{Damgaard:2007xg}
\bibitem{Damgaard:2007xg}
  P.~H.~Damgaard, T.~DeGrand and H.~Fukaya,
  %``Finite-volume Correction to the Pion Decay Constant in the Epsilon-Regime,''
  JHEP {\bf 0712} (2007) 060
  [arXiv:0711.0167 [hep-lat]].
  %%CITATION = ARXIV:0711.0167;%%

\bibitem{Akemann:2008vp}
  G.~Akemann, F.~Basile and L.~Lellouch,
  %``Finite size scaling of meson propagators with isospin chemical potential,''
  JHEP {\bf 0812}, 069 (2008)
  [arXiv:0804.3809 [hep-lat]].

%
\bibitem{Hansen:2011mc}
  M.~T.~Hansen and S.~R.~Sharpe,
  %``Determining low-energy constants in partially quenched Wilson chiral perturbation theory,''
  Phys.\ Rev.\ D {\bf 85} (2012) 054504
  [arXiv:1112.3998 [hep-lat]].


\bibitem{Bar:2003mh}
  O.~Bar, G.~Rupak and N.~Shoresh,
%  "Chiral perturbation theory at O(a**2) for lattice QCD,"
 Phys.\ Rev.\  D {\bf 70} (2004) 034508
  [arXiv:hep-lat/0306021].


\end{thebibliography}
\end{document}